# Orbital dynamics during an ultrafast insulator to metal transition

Sergii Parchenko[1*], Eugenio Paris[1*], Daniel McNally[1], Elsa Abreu[2], Marcus Dantz[1], Elisabeth M. Bothschafter[1], Alexander H. Reid[3], William F. Schlotter[3], Ming-Fu Lin[3], Scott F. Wandel[3], Giacomo Coslovich[3], Sioan Zohar[3], Georgi L. Dakovski[3], Joshua. J. Turner[3], Stefan Moeller[3], Yi Tseng[1], Milan Radovic[1], Conny Saathe[4], Marcus Agaaker[4,5], Joseph E. Nordgren[5], Steven L. Johnson[2], Thorsten Schmitt[1†], and Urs Staub[1‡]

[1] Swiss Light Source, Paul Scherrer Institute, CH-5232 Villigen PSI, Switzerland
[2] Institute for Quantum Electronics, ETH Zürich, 8093 Zürich, Switzerland
[3] LCLS, Stanford National Accelerator Laboratory (SLAC), Menlo Park, California 94025, USA
[4] Lund University, MAX IV Laboratory, Box 118, SE-22100 Lund, Sweden
[5] Uppsala University, Department of Physics & Astronomy, S-75120 Uppsala, Sweden

**Phase transitions driven by ultrashort laser pulses have attracted interest both for understanding the fundamental physics of phase transitions and for potential new data storage or device applications. In many cases these transitions involve transient states that are different from those seen in equilibrium. To understand the microscopic properties of these states, it is useful to develop elementally selective probing techniques that operate in the time domain. Here we show fs-time-resolved measurements of V L-edge Resonant Inelastic X-Ray Scattering (RIXS) from the insulating phase of the Mott-Hubbard material $V_2O_3$ after ultrafast laser excitation. The probed orbital excitations within the d-shell of the V ion show a sub-ps time response, which evolve at later times to a state that appears electronically indistinguishable from the high-temperature metallic state. Our results demonstrate the potential for RIXS spectroscopy to study the ultrafast orbital dynamics in strongly correlated materials.**

As seen in many materials, a temperature-driven insulator to metal transition (IMT) is characterized by a sudden resistivity change upon heating that typically spans several orders of magnitude.[1] IMTs often involve a complex interplay of microscopic interactions including structural, electronic and magnetic degrees of freedom. In many cases, IMT and similar phase transitions can be triggered by an intense pulse of light [2-6]. These light-driven transitions often involve intermediate states that are not present in equilibrium and are visible only using time-resolved methods[7].

The IMT in $V_2O_3$ has been widely studied in equilibrium. [8-10] $V_2O_3$ undergoes a Mott-Hubbard IMT at 160 K, transforming upon heating from an antiferromagnetic insulating phase to a paramagnetic metallic phase.[10] The IMT is accompanied by a structural phase transition from a low temperature (LT) monoclinic phase to a high-temperature (HT) rhombohedral corundum phase (see inset figure 1 for HT structure). The primary mechanism for the IMT is ascribed to electron correlation effects.[11] Although the formal oxidation state of the V ions in $V_2O_3$ is $V^{3+}$ (valence shell configuration V $3d^2$), the actual electronic configuration of the system, as well as the nature of the orbital structure, is still unclear. The orbital occupation has been investigated by polarization-dependent x-ray absorption spectroscopy[12] indicating a high-spin state electron configuration of $3d^2$ for the V ions that have a different orbital occupancy in the different phases. More recently, an LDA-DMFT study[13] predicted that the IMT is in fact driven by correlation-induced enhancement of the crystal-field splitting.

The ultrafast dynamics of laser-excited $V_2O_3$ has been studied using a variety of methods.[14-16] Optical studies found strong changes of the speed of sound (acoustic phonons) through the IMT. More recently, a combination of ultrafast angle-resolved photoemission (ARPES), X-ray diffraction and optical reflectivity experiments in the laser-excited HT phase of $V_2O_3$ revealed a long-lived transient electronic state lasting several picoseconds.[16] This state was found to be associated with an increased dimerization of the V-V ions (see inset Figure 1) occurring around 1 ps after excitation. However, ARPES and optical reflectivity measurements of the time dependent electronic structure do not give direct specification on the occupied and unoccupied orbitals involved in the ultrafast dynamics of laser induced IMTs. A particularly powerful technique to study the elemental and orbital resolved electronic structure is resonant inelastic x-ray scattering (RIXS). Tuning the photon energy to an absorption edge of an atom, thereby initiating an electronic transition between a core level and an unoccupied valence band state, makes RIXS extremely sensitive to electronic excitations within charge, orbital, spin and lattice degrees of freedom. The created energetic disturbance in the intermediate core-excited state is relaxed through a recombination of a valence electron with the core-hole on a femtosecond timescale re-emitting a photon. This leaves the material in an excited state that represents directly an elementary excitation of the electronic, orbital, magnetic or lattice system.[17-20] RIXS has been previously employed to study $VO_2$ [21] and $V_2O_3$ [22], although for the latter case limited energy resolution provided information only on the overall tendencies of the orbital physics. First ultrafast hard x-ray RIXS has been used to study magnetic excitations in $Sr_2IrO_4$.[23]

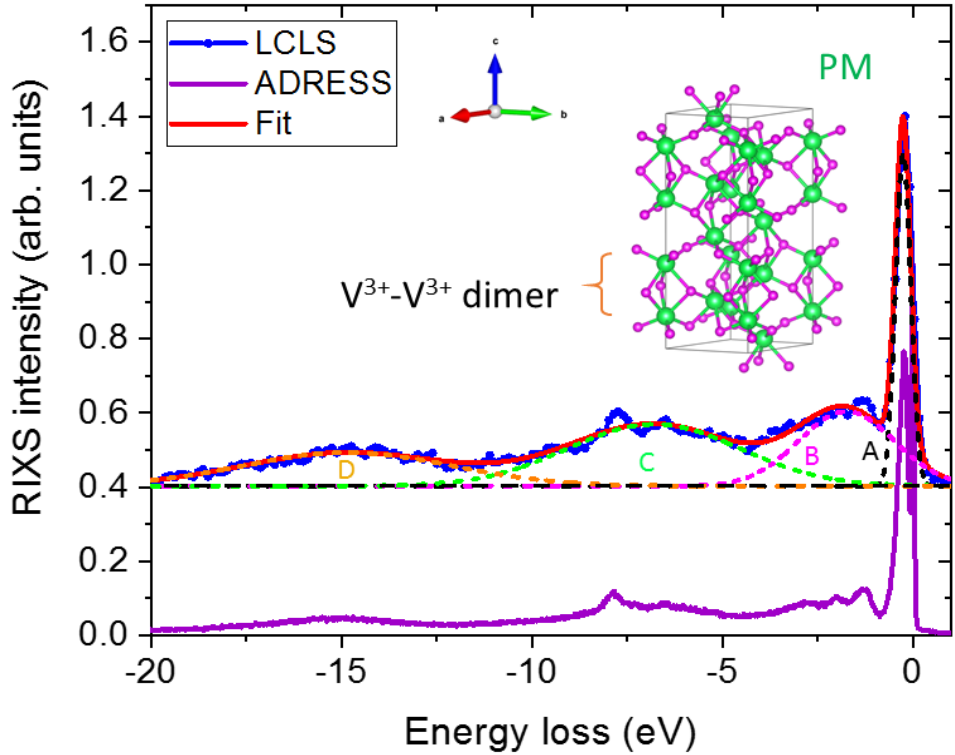

*Figure 1. RIXS spectra of insulating $V_2O_3$ taken at the SXR beamline of the LCLS (blue curve: top) taken at 50K and at the ADRESS beamline of the Swiss Light Source (violet curve: bottom) taken at 20K. Both spectra are normalized to main peak intensity at ~0.2 eV. The red line is a fit to a sum of Gaussian functions representing d-d (A and B) excitations, charge transfer (C) excitations and fluorescence (D) contributions to the RIXS spectrum. The spectra are vertically shifted for better visibility. The dashed curves show the individual components of the fit. The inset shows the crystal structure for the high-temperature phase.*

Here we present a RIXS measurement of the ultrafast response of the d-d excitations in $V_2O_3$ after photoexcitation of the insulating state, with sufficient time and energy resolution to characterize the transient evolution of the V 3d orbitals. We find an ultrafast sub-ps change in the d-d excitations between the different V 3d-$t_{2g}$ orbitals followed by a transient recovery at intermediate (ps) timescale and a slower, quasi-thermal IMT.

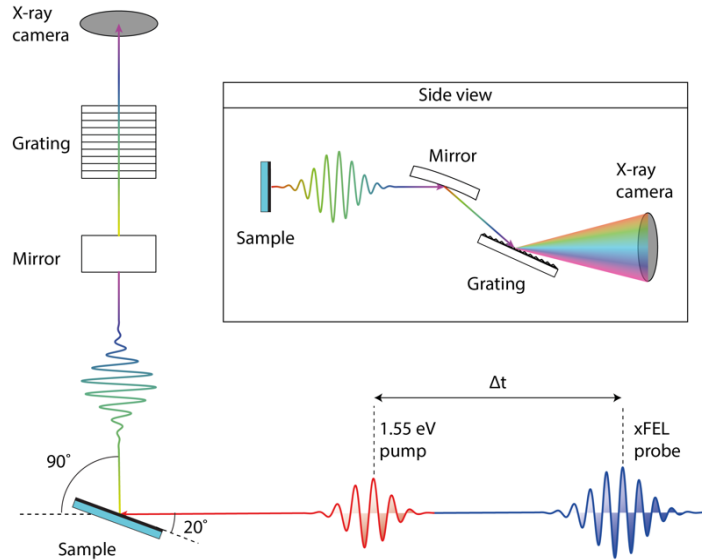

Figure 2 Schematics of the RIXS spectrometer setup at the SXR beamline of the LCLS.

**Results and Discussion**

The experimental configuration is described in Figure 2 in a schematic way. The setup consists of a spectrometer mounted at a 90 degree scattering angle in horizontal geometry. Both the incoming x-ray and optical pulses are close to collinear. The incident π-polarized x-rays reduce the elastic scattering probability due to the polarization dependence of Thomson scattering at the Brewster angle. The VLS grating disperses the scattered x-rays vertically allowing to collect a large solid angle in the horizontal plane employing an extended horizontal line focus, thereby reducing sample damage at the same time as increasing the collected signal. More experimental details can be found in methods.

Figure 1 (a) shows a comparison of a V $L_3$-edge RIXS spectrum in the insulating LT state at a temperature of 20 K recorded at the ADRESS beamline of the Swiss Light Source with a spectrum recorded under similar conditions (except with significant lower energy resolution) at a temperature of 50 K using the RIXS setup at the SXR beamline [24] of the X-ray free-electron-laser LCLS at the SLAC National Accelerator Laboratory. The spectra show an intense peak at 0.2 eV energy loss, followed by several overlapping, less intense peaks located between 1-4 eV. The low energy peaks (A and B) are interpreted to be orbital d-d excitations, as commonly done for transition metal ions [25]. The region between 5-10 eV energy loss is composed of charge transfer (CT) excitations (C), while above approximately 12 eV energy loss the spectrum contains mainly non-Raman fluorescence response of the valence band, as found for vanadium oxides of different oxidation states (D). [26] Both spectra are in good agreement considering the difference in energy resolution between the two data sets. The red line represents a fit to a sum of several Gaussians (see SM) to better quantify changes of the lowest lying d-d excitation (~0.2 eV) in the RIXS spectra.

Figure 3 (a) and (b) show proposed energy diagrams for the antiferromagnetic insulating (AFI) and paramagnetic metallic (PM) phases, respectively. The 3d orbitals of $V^{3+}$ contain two electrons, and in the crystal field this results in doubly occupied $e^{\pi}_g$ orbitals. Our static RIXS data indicate that the first unoccupied orbital ($a_{1g}$) is located at ~0.2 eV above the $e^{\pi}_g$ orbital, followed by $e_{1g}$ orbitals at significantly higher energies. In the AFI state, the crystal structure is monoclinic with more localized electronic states, as compared to the high temperature rhombohedral phase. This effect can also be seen in the d-d excitation, having a narrower $e^{\pi}_g$ to $a_{1g}$ transition (A) at 20K compared to room temperature as seen in Figure 3c inset. A similar effect has recently been observed in nickelates. [27] These equilibrium spectra from ADRESS are convolved with the energy resolution of the time-resolved experiment to better detect possible spectral differences. A Gaussian fit to the d-d excitation of the equilibrium data shows that the total spectral weight of this excitation remains constant across the phase transition within experimental accuracy, although the width and peak height of this lowest lying excitation is strongly affected by the IMT. In the more delocalized PM state, the states are broader representing an overlap of the $a_{1g}$ band with the conduction band – making $V_2O_3$ metallic without a significant change in the energy level position. $VO_2$ showed similar RIXS spectra when passing through its IMT, where the broadening was interpreted as originating from an additional magnetic interaction induced singlet triplet spitting. [21]

Independent of the microscopic origin of the broadening, our time resolved $V_2O_3$ data indicate that the laser excitation does induce the IMT as the static PM RIXS spectra closely resemble the spectra 50 ps after the excitation as can be seen from comparing Figure 3c with its inset. The corresponding changes of the peak heights of the first RIXS excitation between the antiferromagnetic insulating and the paramagnetic metallic state in thermal equilibrium (using the ADRESS data convolved with the experimental resolution of the LCLS RIXS setup) results in very similar ratios $R_{equilibrium} = \frac{I_{20K}}{I_{200K}} = 1.31 \pm 0.02$ compared to the dynamic ones $R_{dynamic} = \frac{I_{-1ps}}{I_{50ps}} = 1.29 \pm 0.06$.

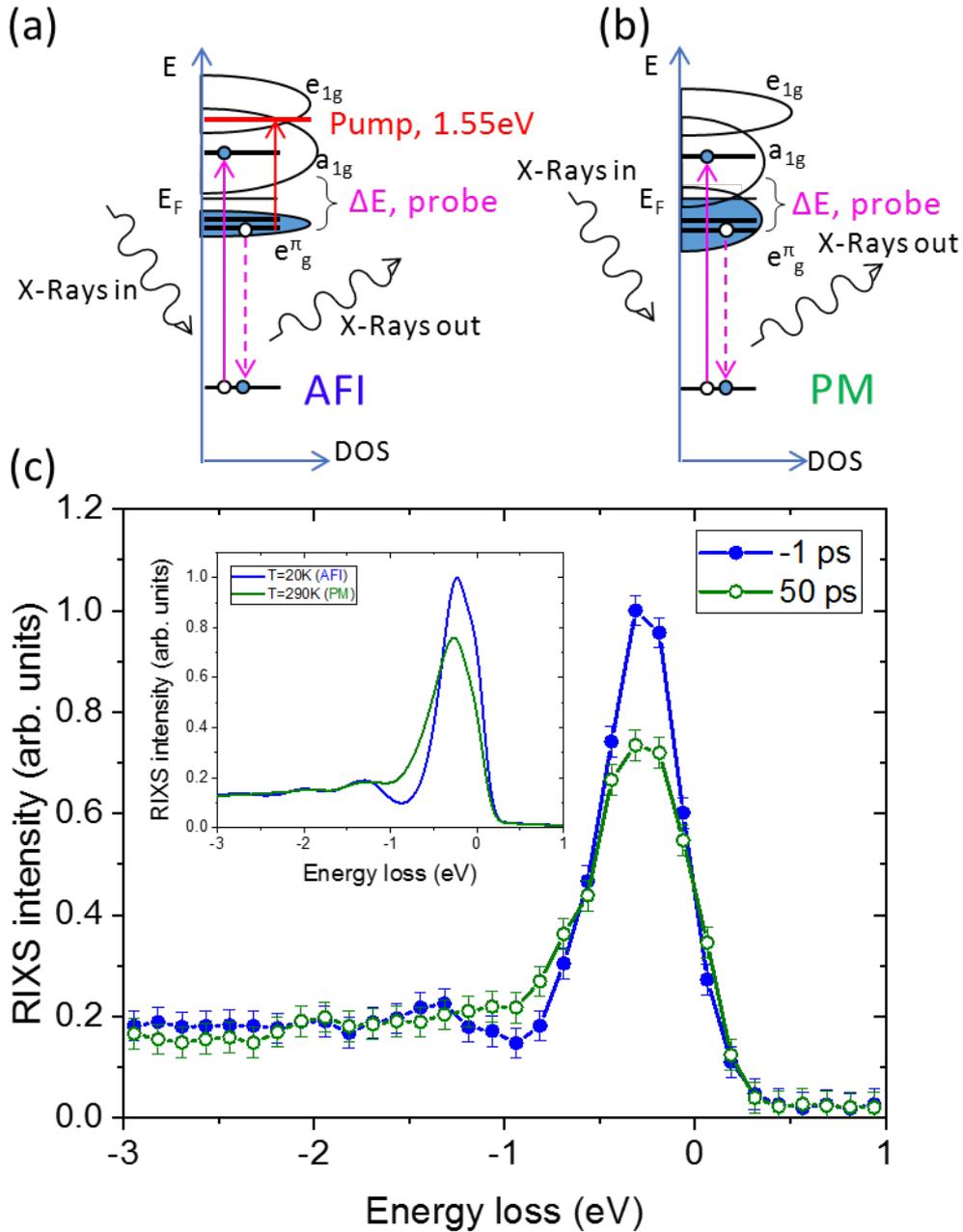

*Figure 3. Schematic energy diagram for (a) AFI and (b) PM phase (50K), visualizing the pumping process and the subsequent RIXS excitations in the two phases. (c) Energy loss RIXS spectra covering the lowest lying d-d excitations before and after photoexcitation with a fluence of Φ=12 mJ/cm$^2$. The inset is showing static RIXS data for AFI (20K) and PM (200K) phases that are convolved with the energy resolution of the time resolved experiment.*

The time scale of this laser-induced transition can be explored using intermediate pump-probe delay times. Figure 4 shows the time evolution of the peak height (a) and width (b) of the 0.2 eV d-d excitation, as obtained from fitting the RIXS data collected with 12 and 24 mJ/cm$^2$ incident laser pump fluences. Within measurement uncertainties, the changes in the peak height and width are complete within 5 ps. For times longer than 5 ps the overall magnitude of these changes is similar for both fluences, although the higher fluence appears to result in a slightly larger change, indicative for an electronically thermalized state (TE). On shorter time scales, we see that the peak height for both fluences drops significantly on a time scale less than one picosecond (photoexcited state (PE)). Optical pump-probe experiments show changes in the optical conductivity over several tens of femtoseconds,[16] suggesting that these changes in the RIXS signal could be much faster than we are able to observe here. This initial change is likely due to the laser-driven redistribution of electrons, resulting in a depopulation of the vanadium 3d valence band as electrons are promoted into the conduction band.

The dynamics measured for the two fluences differs on intermediate time scales, between 1 and 5 ps. While the amplitude of the peak decreases further for the 24 mJ/cm$^2$ data, the data for 12 mJ/cm$^2$ fluence appears to plateau or even slightly recover over a 2 ps time scale. The time scale of this apparent partial recovery has a similar time scale than that of the transient dimerization (DM) of V-V pairs observed in the PM phase by ultrafast x-ray diffraction.[16] This reduced V-V pair distance has been interpreted as a direct consequence of the electronic excitation of the system. [14, 16] Since the V-V dimerization is not strongly influenced by the phase transition, it is plausible to speculate that a similar effect might also occur in the present experiment. Smaller interionic V-V distances from dimerization would increase the Coulomb repulsion, which would lead to an increase in localization of V 3d electrons as in the case for the thermally driven IMT in VO$_2$. [1, 28] This localization partially counteracts the delocalization caused by the metal-insulator transition, resulting in an apparent delay of the drop in the d-d peak. As the system relaxes to local thermal equilibrium, the highly excited electrons transfer heat to the lattice and the electronically-induced V-V dimerization relaxes to its value in thermal equilibrium at an elevated temperature within the PM phase.

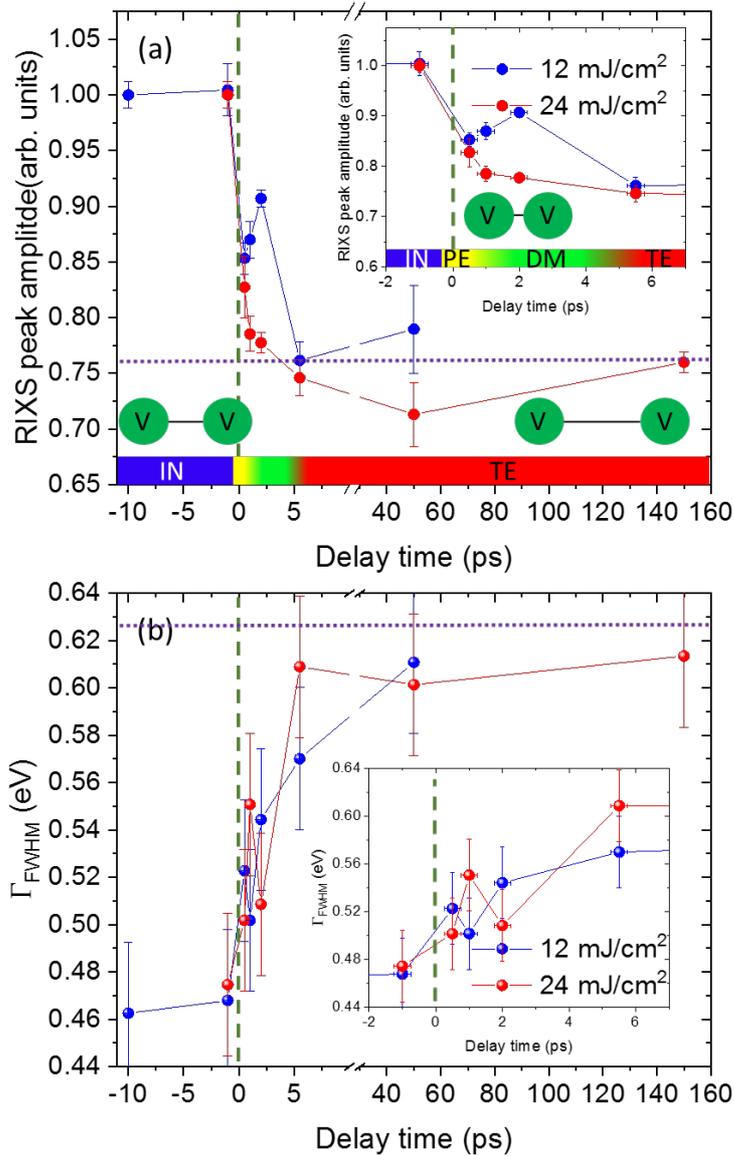

*Figure 4. Time dependence of the RIXS peak amplitude (a) and the full width at half maximum (FWHM) (b) for two excitation fluences. The insets show the dependence over the first few picoseconds after the excitation. The color bar on the bottom of panel (a) refers to different transient states with an initial (IN), an immediate photo excited (PE), a "dimerized" (DM) and a thermalized state (TE).*

Figure 4b shows the changes in the d-d peak width with time. Interestingly, the changes of the width at 24 mJ/cm$^2$ seem to occur on a slower timescale than do the changes in the peak height. This suggests that the integrated intensity changes with time, which is somewhat surprising since it is constant across the thermally driven transition. At short times after laser excitation (between 0 and 3 ps) it indeed appears that the integrated intensity is somewhat depressed (see SFig. 6). This may be a consequence of strong depopulation of the V 3d band at high fluences that reduces the available 3d electrons that can participate in the RIXS process. This might be related to the absence of a plateau in the peak height change at these intermediate times. However, this absence might simply be caused by non-linear response between structural distortion and electronic delocalization as a function of fluence.

**Summary**

In summary, we demonstrate that femtosecond time-resolved RIXS with sufficient energy and time resolution allows us to see intermediate non-thermal states during the photoexcited ultrafast IMT in $V_2O_3$. In particular, we find an apparent delay of the electronic signatures of the phase transition when driven at low fluences, which may be related to a transient dimerization of vanadium ions. Combined ultrafast diffraction and RIXS measurements of the laser induced IMT would be a useful confirmation of this scenario. Together with the recent work pointing out the importance of disorder through an IMT in $VO_2$ [29], these results show that additional microscopic information on the dynamics of the electronic states is required to understand IMT's in strongly correlated electron systems. New detectors that allow for shot-to-shot correction of the time-resolved data will in the future enable soft x-ray RIXS to provide important missing pieces of information on ultrafast electronic and magnetic processes in solids.

**Data Availability**

The datasets generated during and/or analysed during the current study are available from the corresponding authors on reasonable request.

**Acknowledgement**

E.A. acknowledges support from the ETH Zurich Postdoctoral Fellowship Program and from the Marie Curie Actions for People COFUND Program. E.M. B. acknowledges funding from the European Community's Seventh Framework Programme (FP7/2007-2013) under Grant No. 290605 (PSI-FELLOW/COFUND). Financial support of the Swiss National Science Foundation and its National Center of Competence in Research, Molecular Ultrafast Science and Technology (NCCR MUST) is greatly acknowledged. The static synchrotron experiments have been performed at the ADRESS beamline of the Swiss Light Source at the Paul Scherrer Institut (PSI). D.M. was supported by the Swiss National Science Foundation through the NCCR MARVEL. E.P. and Y.Z. were supported by the Sinergia network Mott Physics Beyond the Heisenberg Model (MPBH) (SNSF Research Grant CRSII2_160765/1 and CRSII2_141962). M.D. was partially funded by the Swiss National Science Foundation within the D-A-CH programme (SNSF Research Grant 200021L 141325). Use of the Linac Coherent Light Source (LCLS), SLAC National Accelerator Laboratory, is supported by the U.S. Department of Energy, Office of Science, Office of Basic Energy Sciences under Contract No. DE-AC02-76SF00515.

# Methods

High quality $V_2O_3$ thin films were grown with (001) out-of-plane direction on (0001)-oriented $Al_2O_3$ substrates with pulsed laser deposition. The epitaxial growth conditions are achieved with an oxygen partial pressure of $3 \times 10^{-7}$ mbar while the temperature of the substrate is kept at 750 °C. The $V_2O_3$ polycrystalline target was ablated using a KrF excimer laser ($\lambda = 248$ nm) with 2 Hz repetition rate and a fluence of ~ 1 J/cm$^2$. The obtained thickness was around 100nm. Preliminary characterization to determine the damage threshold of the sample was performed at normal incidence and revealed no visible damage of the film up to an incident fluence of 100 mJ/cm$^2$.

The static V L-edge Resonant Inelastic X-ray Scattering (RIXS) measurements were carried out at the ADRESS beamline of the Swiss Light Source at the Paul Scherrer Institut, Switzerland [30]. The high-brilliance x-ray beam was monochromatized using a collimated-light plane grating monochromator and focused down to a spot of ≤ 4x55 μm$^2$ size at the sample position using an ellipsoidal refocusing mirror. The incoming radiation is linearly polarized, with the polarization vector parallel to the scattering plane. The scattering angle was set to $2\theta = 90°$ while all measurements were performed at specular condition, i.e. with the incoming beam impinging at an angle of approximately 45° with respect to the sample surface. The combined energy resolution was 55 meV FWHM, determined by collecting the elastic scattering from a carbon tape reference.

The time resolved experiments were performed at the LCLS free electron laser using the resonant inelastic scattering spectrometer recently established at SXR beamline. [24] Fig. 2 shows the scheme of the experiment. We used 800 nm p-polarized ultrashort laser pulses as a pump and 120 fs π-polarized x-ray pulses as a probe, with an energy corresponding to the pre-edge of the vanadium $L_2$ absorption, as indicated by the arrow in SFigure 3. Pump and probe beams propagate collinearly with 20° incidence angle from the sample surface and were focused to 5370x1330 μm$^2$ and 2050x50 μm$^2$ spot size for the pump and probe beams, respectively. The x-ray beam was monochromatized using a plane grating monochromator and focused on the sample using KB-optics. The scattering angle was set to 90°. We estimate the combined energy resolution to be 200 meV FWHM, obtained by operating the beamline with 50 μm exit slits and with the spectrometer grating set to the 2$^{nd}$ diffraction order. We used an Andor Newton DO940P charge coupled device (CCD) camera to collect the RIXS signal. The camera was continuously exposed for 10 min for the different time delays, repeating this process several times. During the acquisition, drifts in the pump-probe delay were corrected by manually adjusting online traces of the x-ray and laser arrival times using the shot-to-shot time arrival monitoring system to better than 0.4 ps (see also SM). To avoid artifacts due to the cosmic background radiation within the acquisition time, a threshold-based algorithm was implemented to filter the signal (see SM). The energy calibration was performed by recording the $L_\alpha$ and $L_\beta$ emission lines of a reference Zn foil, using the 4$^{th}$ order of the spectrometer grating.

The main parameters (RIXS amplitude and FWHM of the lowest energy d-d peak), which represent the dynamics of the laser induced IMT was extracted from fitting the spectra to four Gaussian functions. The parameters for the higher lying d-d excitation, the charge transfer excitation and the fluorescence contribution were kept fixed for simplicity to the values obtained for the data taken before $t_0$.

\* Authors with equal contributions

† thorsten.schmitt@psi.ch

‡ <u>urs.staub@psi.ch</u>

**AUTHOR CONTRIBUTIONS**

U.S. and T. S conceived and coordinated the study. E.P., D. M., E. A., M. D., E. M. B., A. H. R., W. F. S., M-F. L., G. L. D., J. J. T., S. M., C. S., M. A., J. E. N., S. L. J, T. S., and U. S. performed the ultrafast RIXS experiments, G. G. and S.F.W. were responsible for the laser setup, S. Z. helped with the data acquisition and the treatment, S. P. and E. P. analyzed the data, E. A. determined the laser damage threshold of the sample, D. M., E. P. Y. T. and T.S. performed the static RIXS characterization, D. M, E. P. and M. R. have grown and characterized the thin films, S.P., E. P., T. S. and U. S. wrote the manuscript with input from all authors.

**ADDITIONAL INFORMATION**

Supplementary information available online (url to be inserted during production)

Competing interests: The authors declare no competing Financial or Non-Financial interests.